\documentclass[onecolumn,letterpaper,11pt,draftclsnofoot]{IEEEtran}
\usepackage{epsfig,subfigure,cite,amssymb}

\newtheorem{theorem}{Theorem}

\begin{document}

\title{\LARGE{Lifetime Improvement in Wireless Sensor Networks via Collaborative Beamforming and Cooperative Transmission}}
\author{Zhu Han$^*$ and H. Vincent Poor$^+$\\
$^*$Department of Electrical and Computer Engineering, Boise State
University, Idaho, USA\\
$^+$Department of Electrical Engineering, Princeton University,
New Jersey, USA \thanks{This research was supported by the
National Science Foundation under Grants ANI-03-38807,
CCR-02-05214, and CNS-06-25637.}}

\maketitle

\begin{abstract}\thispagestyle{empty}

Collaborative beamforming (CB) and cooperative transmission (CT)
have recently emerged as communication techniques that can make
effective use of collaborative/cooperative nodes to create a
virtual multiple-input/multiple-output (MIMO) system. Extending
the lifetime of networks composed of battery-operated nodes is a
key issue in the design and operation of wireless sensor networks.
This paper considers the effects on network lifetime of allowing
closely located nodes to use CB/CT to reduce the load or even to
avoid packet-forwarding requests to nodes that have critical
battery life. First, the effectiveness of CB/CT in improving the
signal strength at a faraway destination using energy in nearby
nodes is studied. Then, the performance improvement obtained by
this technique is analyzed for a special 2D disk case.  Further,
for general networks in which information-generation rates are
fixed, a new routing problem is formulated as a linear programming
problem, while for other general networks, the cost for routing is
dynamically adjusted according to the amount of energy remaining
and the effectiveness of CB/CT. From the analysis and the
simulation results, it is seen that the proposed method can reduce
the payloads of energy-depleting nodes by about 90\% in the
special case network considered and improve the lifetimes of
general networks by about 10\%, compared with existing techniques.
\end{abstract}

\newpage
\section{Introduction}\label{sec:intro}\setlength{\baselineskip}{24pt}

In wireless sensor networks \cite{Akyildiz1}, extending the
lifetimes of battery-operated devices is considered a key design
issue that increases the capability of uninterrupted information
exchange and alleviates the burden of replenishing batteries. A
survey of energy constraints for sensor networks in presented in
\cite{Ephremides}. In the literature of sensor networks, there are
many lifetime definitions: the time until the first node failure
\cite{Raghavendra}, \cite{Tassiulas}, \cite{Vainstein}; the time
at which only a certain fraction of nodes survives
\cite{Ephremides} \cite{Wattenhofer}; the time until the first
loss of coverage \cite{Bhardwaj};  mean expiration time
\cite{Toh}; and lifetime in terms of alive flows \cite{Zhang} and
packet delivery rates \cite{Morris}.

To improve network lifetimes, many approaches have been proposed.
In \cite{Tassiulas}, a data routing algorithm has been proposed
with an aim to maximize the minimum lifetime over all nodes in
wireless sensor networks. Upper bounds on lifetime have been
derived in \cite{Bhardwaj} using an energy-efficient tree with
static and dynamic optimization. In \cite{Yates1}, the network
lifetime has been maximized by employing an accumulative broadcast
strategy. The work in \cite{Midkiff1} has considered provisioning
additional energy on existing nodes and deploying relays to extend
the lifetime.

Recently, collaborative beamforming (CB) \cite{CB} and cooperative
transmission (CT) \cite{bib:Aazhang1}\cite{bib:Laneman2} have been
proposed as new communication techniques that fully utilize
spatial diversity and multiuser diversity. The basic idea is to
allow nodes in the network to help transmit/relay information for
each other, so that collaborative/cooperative nodes create a
virtual multiple-input/multiple-output (MIMO) system. Most
existing works on CB/CT concentrate on how to improve performance
at the physical layer. Recently, some works have also studied the
impacts of CB/CT on the design of higher layer protocols. In
\cite{Luo1}\cite{bib:zhu_Ahmed}\cite{Zhu1}, the authors have
investigated the choice of appropriate nodes for cooperation and
how much cooperation should be facilitated. A game theoretic
approach for relay selection has been proposed in
\cite{globecom_zhu}. In \cite{bib:Madsen}, cooperative routing
protocols have been constructed based on non-cooperative routes.

In this paper, we investigate new routing protocols to improve
lifetimes of wireless sensor networks using CB/CT communication
techniques. First, we study the fact that CB/CT can effectively
increase the signal strength at the destination, which in turn can
increase the transmission range. We obtain a closed-form analysis
of the effectiveness of CT similar to that of CB in \cite{CB}.
Then, we formulate the problem so as to maximize the network
lifetime defined as the time at which the first node fails. The
{\em new idea} is that closely located nodes can use CB/CT to
reduce the loads or even avoid packet forwarding requests to nodes
with critical battery lives. From the analysis of a 2D disk case
using CB/CT, we investigate how to bypass the energy-depleting
nodes. Then we propose two algorithms for general networks. If the
information-generation rates are fixed, we can formulate the
problem as a linear programming problem. Otherwise, we propose a
heuristic algorithm to dynamically update the cost for routing
tables according to the remaining energy and willingness for
collaboration. From the analysis and simulation results, we see
that the proposed new routing schemes can reduce the payloads of
the energy-depleting nodes by about 90\% in the 2D disk network
compared to the pure packet-forwarding scheme, and can improve the
lifetimes of general networks by about 10\%, compared to the
schemes in \cite{Tassiulas}.

This paper is organized as follows: In Section \ref{sec:model}, we
describe the system model and study the ability to enhance the
destination signal strength by CB/CT. In Section
\ref{sec:prob_def}, we formulate the lifetime maximization
problem, analyze a 2D disk case, and propose algorithms for
general networks. Simulation results are given in Section
\ref{sec:simulation} and conclusions are drawn in Section
\ref{sec:conclusion}.

\section{System Model and Effectiveness of CB/CT}\label{sec:model}

We consider a wireless network where each sensor node is equipped
with a single ideal isotropic antenna. There is no power control
for each node; i.e., the node transmits with power either $P$ or
$0$. We assume sufficiently separated nodes so that any mutual
coupling effects among the antennas of different nodes are
negligible. For CB, there is no reflection or scattering of the
signal. Thus, there is no multipath fading or shadowing. For CT,
we assume Rayleigh fading.

For traditional direct transmission, if a node tries to reach
another node at a distance of $A$, the signal to noise ratio (SNR)
is given by

\begin{equation}
\Gamma=\frac {PC_0A^{-\alpha}|h|^2}{\sigma ^2},
\end{equation}
where $C_0$ is a constant, $\alpha$ is the propagation loss
factor, $h$ is the channel fading gain, and $\sigma^2$ is the
thermal noise level. Here, $C_0$ includes other effects such as
the antenna gains and is assumed to be unity without loss of
generality. We define the energy cost of such a transmission for
each packet as one unit. We further define $A_0$ to be the maximal
distance over which a minimal link quality $\gamma_0$ can be
maintained, i.e. $\Gamma(A_0)=\gamma_0$.

\begin{figure}[htbp]
\begin{center}
    \epsfig{file=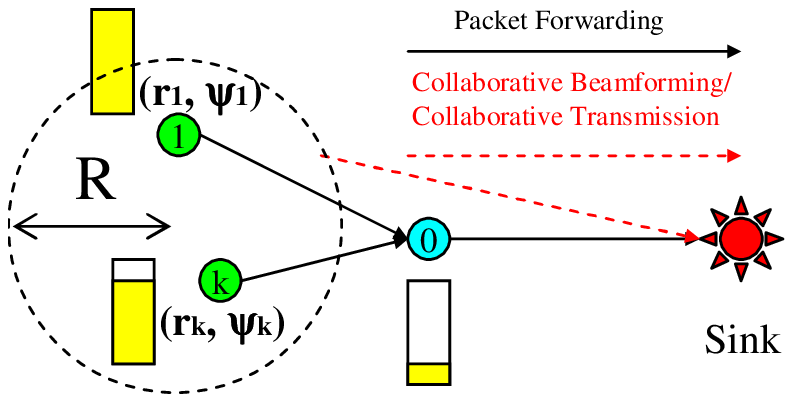,width=100truemm}
\end{center}
\caption{System Model}\label{system_model}
\end{figure}

In Figure \ref{system_model}, we show the system model with CB/CT.
In traditional sensor networks, the only choice a node has is to
forward packets toward the sink which serves as a data gathering
point. This packet forwarding will deplete the energy of the nodes
near the sink, since they have to transmit a lot of other nodes'
packets. To overcome the above problem, we propose another choice
for a node by forming CB/CT with the nearby nodes so as to
transmit further towards the sink. By using the nearby nodes as
virtual antennas of a MIMO system, we can leverage the energy
usage of the nodes having different locations and different
remaining energies. As a result, the network lifetime can be
improved. In the rest of this section, we study how CB and CT can
improve link quality. Notice that CB and CT technologies are
investigated independently in the rest of this paper for lifetime
improvement over wireless sensor networks.

\subsection{Effectiveness of Collaborative Beamforming}

We assume sensor nodes are uniformly distributed with a density of
$\rho$. If there are a total of $N$ nodes for collaborative
beamforming within a disk with radius $R$. We have
\begin{equation}
N=\lfloor \rho \pi R^2\rfloor.\label{N_R}
\end{equation}
Each node has polar coordinate $(r_k,\psi_k)$ with the origin at
the disk center. The distance from the center to the beamforming
destination is $A$. The Euclidean distance between the $k^{th}$
node and the beamforming destination can be written as:
\begin{equation}
d_k=\sqrt{A^2+r_k^2-2r_kA\cos(\phi-\psi_k)},
\end{equation}
where $\phi$ denotes azimuthal direction and is assumed to be a
constant. By using loop control or the Global Positioning System,
the initial phase of node $k$ can be set to
\begin{equation}
\psi _k=-\frac {2\pi}{\lambda} d_k(\phi),
\end{equation}
where $\lambda$ is the wavelength of the radio frequency carrier.

Define $\textbf z=[z_1\dots z_N]^T$ with
\begin{equation}
z_k=\frac {r_k} R \sin ( \psi_k-\phi/2).
\end{equation}
The array factor of CB can be written as
\begin{equation}
F(\phi|\textbf z)=\frac 1 N \sum_{k=1}^N e^{-j4\pi R \sin(\frac
\phi 2)z_k/\lambda}.
\end{equation}
The far-field beam pattern can be defined as:
\begin{equation}
P(\phi|\textbf z)=|F(\phi|\textbf z)|^2 =\frac 1 N +\frac 1{N^2}
\sum_{k=1}^N e^{-ja(\phi)z_k} \sum_{l\neq k} e^{ja(\phi)z_l},
\end{equation}
where
\begin{equation}
a(\phi)=\frac {4\pi R \sin \frac \phi 2}{\lambda}.
\end{equation}

Define the directional gain $D_{av}^{CB}$ as the ratio of radiated
concentrated energy in the desired direction over that of a single
isotropic antenna. From Theorem 1 in \cite{CB}, for large $\frac R
\lambda$ and $N$, the following lower bound for far-field
beamforming is tightly held:
\begin{equation}
    \frac{D_{av}^{CB}} N \geq \frac 1 {1+\mu\frac
    {N\lambda}{R}},\label{bound}
\end{equation}
where $\mu\approx 0.09332$.

Considering this directional gain, we can improve the direct
transmission by a factor of $D_{av}^{CB}$. Consequently, the
transmission distance is improved. Notice that this transmission
distance gain for one transmission is obtained at the expense of
consuming a total energy of $N$ units from the nearby nodes.

\subsection{Effectiveness of Cooperative Transmission}

In this subsection, we discuss another technique for improving the
link quality. Similar to the CB case, we assume $N$ nodes are
uniformly distributed over a disk of radius $R$. The probability
density function of the node's distance $r$ from the center of the
disk is given by
\begin{equation}
q(r)=\frac {2r} {NR^2}, \ \ 0\leq r \leq R,
\end{equation}
and the node's angular coordinate $\psi$ is uniformly distributed
on $[0,2\pi)$.

Suppose at stage $1$, node $1$ transmits to the next hop or sink
which is called the destination. The received signals at node $2$
through node $N$ at stage $1$ can be expressed as:
\begin{equation}
z_k=\sqrt{P r_k^{-\alpha}}h_k^r x+n_k^r,\ k=2,\dots N,
\end{equation}
and the received signal at the sink at stage $1$ is given by
\begin{equation}
y_1=\sqrt{P d_1^{-\alpha}}h_1 x+n_1.
\end{equation}
Then in the following stages, node $2$ through node $N$ relay node
$1$'s information if they decode it correctly. The received
signals at the destination in the subsequent stages are
\begin{equation}
y_k=\sqrt{P d_k^{-\alpha}}h_k x+n_k,\ k=2,\dots N.
\end{equation}
In the rest of this subsection, we call node $1$ the source and
nodes $2$ through $N$ the relays. Here, $P$ is the transmitted
power. $h_k^r$ and $h_k$ are the channel fading gains of
source-relay and relay-destination, respectively. The channel
fading gains are modeled as independent, circularly-symmetric,
complex, Gaussian random variables with zero mean and unit
variance. $x$ is the transmitted data symbol having unit energy.
$n_k$ and $n_k^r$ are independent thermal noises at the
destination and relay for node $k$, respectively. Without loss of
generality, we assume that all noises have variance $\sigma ^2$.

\begin{theorem}
Define $D_{av}^{CT}$ to be the number of times a link can be
enhanced at the destination using CT. Under the far-field
condition and the assumption that channel links between the source
and the relays are sufficiently good, we have the following
approximation:
\begin{equation}\label{approx_CC}
\frac {D_{av}^{CT}}{N}\approx \frac {1+(N-1)\ _2 F_1\left(\frac 2
\alpha, -L; \frac {\alpha+2}\alpha ; \frac
{\sigma^2R^\alpha}{4P}\right)}{N},
\end{equation}
where $L$ is the frame length and $_2 F_1$ is the hypergeometric
function
\begin{equation}
_2 F _1 (a,b;c;z)=\sum_{n=0}^{\infty} \frac {(a)_n(b)_n}{(c)_n}
\frac {z^n}{n!},
\end{equation}
and where $(a)_n$ is the Pochhammer symbol defined as
\[
(a)_n=a(a+1)\cdots(a+n-1).
\]
The calculation of hypergeometric functions is discussed in
\cite{functions}.
\end{theorem}

\begin{proof}
The received SNR at the $k^{th}$ node at stage $1$ can be written
as
\begin{equation}
\Gamma_k=\frac {PC_0r_k^{-\alpha}|h_k^r|^2}{\sigma ^2},
\end{equation}
where $|h_k^r|^2$ is the square magnitude of the channel fade and
follows an exponential distribution with unit mean.

Without loss of generality, we assume BPSK modulation is used and
$C_0=1$. The probability of successful transmission of a packet of
length $L$ is given by
\begin{equation}\label{P_suc}
P_r^k(r_k)=\left( \frac 1 2 + \frac 1 2 \sqrt {\frac {P}{P+\sigma
^2r^{\alpha}_k}}\right) ^L.
\end{equation}
For fixed $(r_k,\psi_k)$, the average power that arrives at the
destination can be written as
\begin{equation}
D^{CT}=\sum_{k=1}^N P d_k^{-\alpha}P_r^k.
\end{equation}
For node $1$, $r_1=0$. We can write the average gain in link
quality in the following generalized form:
\begin{equation}\label{D_c_av_1}
D^{CT}_{av}=\sum_{k=1}^N \int _0 ^R \int _0 ^{2\pi} \frac
{A^\alpha} {2\pi} d_k^{-\alpha} P_r^k (r_k) q(r_k) dr_k d\psi _k.
\end{equation}
Since each node is independent of the others, we omit the
designator $k$ and rewrite (\ref{D_c_av_1}) as
\begin{equation}
D^{CT}_{av}=1+(N-1) \int _0 ^{2\pi} \int _0^R \frac {r} {\pi
A^{-\alpha}R^2} (A^2+r^2-2rA\cos\psi )^{-\frac \alpha 2} P_r(r) dr
d\psi  ,
\end{equation}
where the $1$ on the right-hand side (RHS) is obtained because the
first node's location is at $r_1=0$.  With the far field
assumption, we have
\begin{equation}
\frac 1 {2\pi}\int _0 ^{2\pi} (A^2+r^2-2rA\cos \psi)^{-\frac
\alpha 2} d\psi \approx A^{-\alpha}.
\end{equation}
The average link quality gain is approximated by
\begin{equation}
D^{CT}_{av}\approx 1+(N-1) \frac {2} {R^2} \int _0^R r P_r(r) dr .
\end{equation}
With the assumption of sufficiently good channels between the
source and the relays, using a Taylor expansion, we have the
following approximation of (\ref{P_suc}):
\begin{equation}
P_r(r)\approx (1-\frac {\sigma^2 r ^\alpha}{4P})^L.
\end{equation}

\noindent Since
\begin{equation}
\int _0^R r (1-\frac {\sigma^2 r ^\alpha}{4P})^L dr =\frac 1 2
R^2\ _2 F_1\left(\frac 2 \alpha, -L; \frac {\alpha+2}\alpha ;
\frac {\sigma^2R^\alpha}{4P}\right),
\end{equation}
we obtain (\ref{approx_CC}).
\end{proof}

\begin{figure}[htbp]
\begin{center}
    \epsfig{file=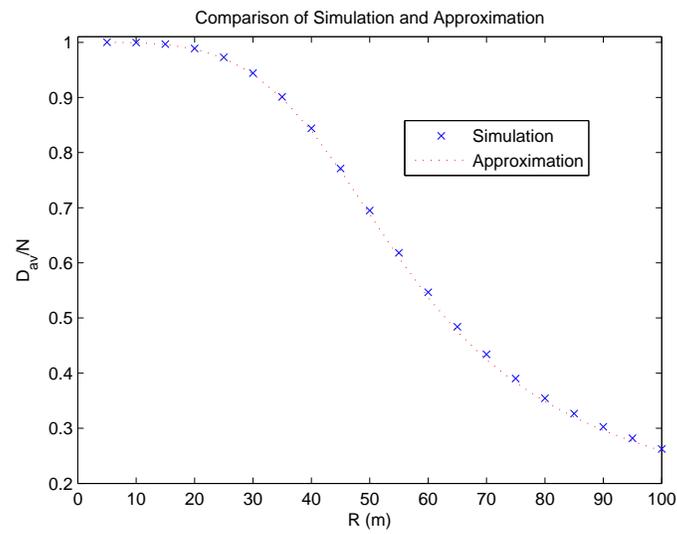,width=100truemm}
\end{center}
\caption{Approximation of CT Effectiveness: Signal Enhancement vs.
Radius}\label{CC_compare}
\end{figure}

In Figure \ref{CC_compare}, we compare the numerical and
analytical results of $D_{av}$ for different radii $R$. Here
$A=1000$m, $P=10$dbm, $\sigma^2=-70$dbm, $\alpha=4$, $L=100$, and
$N=10$. When $R$ is sufficiently small; i.e., when the cooperative
nodes are close to each other, the average link gain over the
number of cooperative nodes is almost one. When $R$ increases, the
efficiency of transmission to a faraway destination decreases
because the links from the source to the relays degrade. We can
see that the numerical results fit the analysis very well, which
suggests that the approximation in (\ref{approx_CC}) is a good
one.

\section{CB/CT Lifetime Maximization\label{sec:prob_def}}

In this section, we first define the lifetime of sensor networks
and formulate the corresponding optimization problem. Then, using
a 2D disk case, we demonstrate the effectiveness of lifetime
improvement analytically using CB/CT. Finally, two algorithms are
proposed for general network configurations with fixed and dynamic
information rates, respectively.

\subsection{Problem Formulation}

\begin{figure}[htbp]
\begin{center}
    \epsfig{file=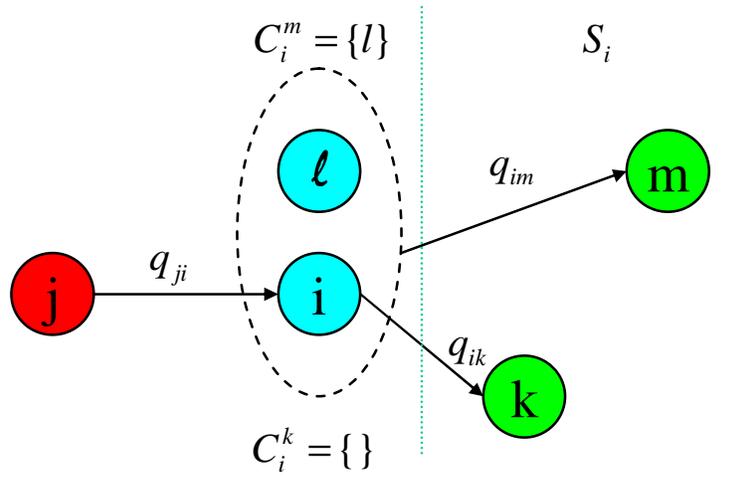,width=100truemm}
\end{center}
\caption{CB/CT Routing Model}\label{routing_mod}
\end{figure}

In Figure \ref{routing_mod}, we show the routing model with CB/CT.
A wireless sensor network can be modeled as a directed graph
$G(M,\mathbb{A})$, where $M$ is the set of all nodes and
$\mathbb{A}$ is the set of all links $(i,j), i,j\in M$. Here the
link can be either a direct transmission link or a link with
CB/CT. Let $S_i$ denote the set of nodes that the $i^{th}$ node
can reach using direct transmission. Denote by $C_i^m$ the set of
nodes that node $i$ needs to use CB/CT with in order to reach node
$m$. In the example in Figure \ref{routing_mod}, $C_i^m=\{ l\}$
and $C_i^k=\{ \}$. A set of origin nodes $O$ where information is
generated at node $i$ with rate $Q_i$ can be written as
\begin{equation}
O=\{ i | Q_i>0,\ i\in M\}.
\end{equation}
A set of destination nodes is defined as $D$ where
\begin{equation}
D=\{ i | Q_i<0,\ i\in M\}.
\end{equation}
Here the sign of $Q_i$ represents the direction of flow.

Let $\textbf q=\{q_{ij}\}$ represent the routing and the
transmission rates. In this paper, we use the lifetime until the
first node failure as an example. Other definitions of lifetime
can be treated similarly. Suppose node $i$ has remaining energy
$E_i$. Then, the lifetime for node $i$ can be written as:
\begin{equation}
T_i(\textbf q)=\frac {E_i}{\sum_{j\in S_i} q_{ij}+\sum _{i\in
C_j^m,\forall j,m} q_{jm}},
\end{equation}
where the first term in the denominator is for direct transmission
and the second term in the denominator is for CB/CT. Notice that
$C_j^m$ is not a function of $\textbf q$, so that we can formulate
the lifetime maximization problem as
\begin{equation}\label{Prob_def}
\max_{\textbf q}\ \min T_i
\end{equation}
\[
\mbox {s.t. }\left\{
\begin{array}{l}
q_{ij}\geq 0, \forall i,j,\\
\sum_{j,i\in S_j} q_{ij} + Q_i=\sum_{k\in S_i}q_{ik},\\
\end{array}
\right.
\]
where the second constraint is for flow conservation. Notice that
there are two differences between (\ref{Prob_def}) and lifetime
maximization in traditional sensor networks without CB/CT. First,
in addition to packet forwarding, the nodes in (\ref{Prob_def})
need to spend additional energy to form CB/CT with other nodes.
Second, there are more available routes for nodes in the formation
of (\ref{Prob_def}) than those in the formulation without CB/CT.

\subsection{2D Disk Case Analysis}

In this subsection, we study a network confined to a 2D disk.
Nodes with equal energy remaining are uniformly located within a
circle of radius $B_0$. One sink is located at the center of the
disk $(0,0)$. Each node has a unit amount of information to
transmit. Here we assume the node density is large enough, so that
each node can find enough nearby nodes to form CB/CT to reach the
faraway destination.

First, we consider the traditional packet-forwarding scheme
without CB/CT. Suppose the node density is $\rho$. The number of
nodes at approximate distance $B$ to the sink is $\rho 2\pi B
\Delta B$, where $\Delta B$ denotes the degree of approximation.
Recall that $A_0$ is the maximal distance for direct transmission.
The nodes located at distance $B+A_0$ transmit to the nodes at
distance $B$, and there are $\rho 2\pi (B+A_0) \Delta B$ such
nodes. We assume a unit information-generation rate for each node.
The number of packets needing transmission for each node at
distance $B$ to the sink is given by
\begin{equation}
N_{pf}(B)=\sum_{n=0}^{\lfloor\frac {B_0-B}{A_0}\rfloor}
\left(1+\frac {nA_0}{B}\right).
\end{equation}
Notice that the burden of forwarding data increases when the nodes
are close to the sink.

If all nodes use their neighbor nodes to communicate with the sink
directly, we call this scheme pure CB/CT. To achieve the range of
$B$, we need $N_{CB/CT}(B)$ nodes for CB/CT so that a direct link
to the sink can be established; i.e.,
\begin{equation}
D_{av}^{CB/CT}\left(N_{CB/CT}(B)\right)=\left(\max(\frac B
{A_0},1)\right)^\alpha.
\end{equation}
For collaborative beamforming, we can calculate
\begin{equation}
N_{CB}(B)\geq \frac 1 2
\left(c_0(2+c_0c_1^2)+c_0^{1.5}c_1\sqrt{4+c_0c_1^2}\right)
\end{equation}
where $c_0=(\max(\frac B {A_0},1))^\alpha$ and $c_1=\mu \lambda
\sqrt{\rho \pi}$. For cooperative transmission, numerical results
need to be used to obtain the inverse of $_2F_1$ in Theorem 1.

\begin{figure}[htbp]
\begin{center}
    \epsfig{file=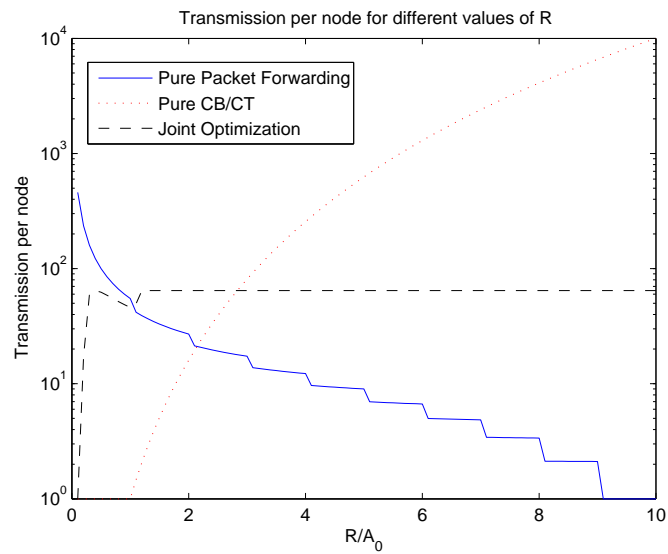,width=100truemm}
\end{center}
\caption{Analytical Results for 2D Disk Case}\label{analysis_fig}
\end{figure}

In Figure \ref{analysis_fig}, we show the average number of
transmissions per node vs. disk size $B_0$. We can see that for
traditional packet forwarding, the node closest to the sink has
the most transmissions per node; i.e., it has the lowest lifetime
if the initial energy is the same for all nodes. On the other
hand, for the pure CB/CT scheme, more nodes need to transmit to
reach the sink directly when $B_0$ is larger. The transmission is
less efficient than packet forwarding, since the propagation loss
factor $\alpha$ is larger than $1$ in most scenarios. The above
facts motivate the joint optimization case in which nodes transmit
packets with different probabilities to emulate traditional packet
forwarding and CB/CT.

As we have stated, if the faraway nodes can form CB/CT to transmit
directly to the sink and bypass these energy-depleting nodes, the
overall network lifetime can be improved. Notice that in this
special case, if the faraway nodes form CB/CT to transmit to nodes
other than the sink, the lifetime will not be improved. For each
node with distance $B$ to the sink, we have the probability
$P_r(B)$ of using CB/CT as

\begin{equation}\label{N_joint}
N_{joint}(B)= [1-P_r(B)+N_{CB/CT}(B)P_r(B)] \left\{
\sum_{n=0}^{\lfloor\frac {B_0-B}{A_0}\rfloor}\left(1+\frac {
nA_0}{B}\right)\Pi _{j=1}^n[1-P_r(B+jA_0)]\right\},
\end{equation}
where the first term on the RHS is the necessary transmitting
energy per packet, and the second term is the total number of
transmitting packets. The goal is to adjust $P_r(B)$ such that the
heaviest payload is minimized; i.e.,
\begin{equation}\label{bisection_prob}
\min_{1 \geq P_r(B)\geq 0} \max N_{joint}(B).
\end{equation}
If $P_r(B)=0,\forall B$ the scheme is traditional packet
forwarding, and if $P_r(B)=1,\forall B>A_0$ the scheme is pure
CB/CT.

Notice that $N_{joint}\geq 1$, and in (\ref{N_joint}) the second
term on the RHS depends on the probabilities of CB/CT being larger
than $B$. Therefore, we can develop an efficient bisection search
method to calculate (\ref{bisection_prob}). We define a
temperature $\kappa$ that is assumed to be equal to or greater
than $N_{joint}(B),\forall B$. We can first calculate $P_r(B_0)$
at the boundary of the network where $N_{joint}(B_0)=\kappa$ and
the second term on the RHS of (\ref{N_joint}) is one. Then, we can
derive all $N_{joint}(B)$ and $P_r(B)$ by successively reducing
$B$. If $\kappa$ is too large, most information is transmitted by
CB/CT, and the nodes faraway from the sink waste too much energy
for CB/CT. In this case, we can reduce $\kappa$. On the other
hand, if $\kappa$ is too small, the nodes close to the sink must
forward too many packets and the resulting $N_{joint}(B)$ is
larger than $\kappa$ when $B$ is small. A bisection search method
can find the optimal values of $\kappa$, $N_{joint}(B)$, and
$P_r(B)$.

\begin{table}
\caption{Payload Reduction vs. Disk Size}
\begin{center}
\begin{tabular}{|c|c|c|c|c|c|}
  \hline
  $B_0$ & 2 & 4 & 6 & 8 & 10 \\
  \hline
  $N_{joint}(B)$ & 2.8 & 10.3 & 23.4 & 42.5 & 64.5 \\

  \hline
  $N_{pf}(B)$ & 52 & 153 & 256 & 358 & 460 \\

  \hline

  Saving \% & 94.6 & 93.3 & 90.9 & 88.1 & 86.0 \\
  \hline
\end{tabular}
\end{center}\label{ana_table}
\end{table}

In Figure \ref{analysis_fig}, we show the joint optimization case
where the node density is sufficiently large. We can see that to
reduce the packet-forwarding burdens of the nodes near the sink,
the faraway nodes form CB/CT to transmit to the sink directly. The
formulation of CB/CT will increase the number of transmissions per
node for themselves, but reduce the number of transmissions per
node for the nodes near the sink. By leveraging between the energy
consumed for CB/CT and packet forwarding, the proposed scheme can
reduce the payload burdens and consequently improve the network
lifetime significantly. In Table \ref{ana_table}, we show the
optimal $N_{joint}(B)$, $N_{pf}(B)$, and the payload reductions
for the proposed joint scheme over traditional packet forwarding.
We can see that the payloads are reduced by about 90\%. The
performance improvement decreases when the size of the network
increases. This is because CB/CT costs too much transmission
energy to connect to the sink directly if the propagation loss is
greater than $1$.

\subsection{General Case Algorithms\label{sec:algorithm}}

In this section, we first consider the case in which the
information-generation rates are fixed for all nodes and develop a
linear programming method to calculate the routing table. Define
$\hat q_{ij}=Tq_{ij}$, where $T$ is the lifetime. If $C_j^m,
\forall j,m$ is known, the problem in (\ref{Prob_def}) can be
written as a linear programming problem:
\begin{equation}
\max \ T \label{linear_prog}
\end{equation}
\[
\mbox{s.t. } \left\{
\begin{array}{l}
\hat q_{ij}\geq 0, \forall i,j;\\
(\sum_{j\in S_i}\hat q_{ij}+\sum _{i\in C_j^m,\forall j,m}\hat q_{jm})\leq E_i, \ \forall i;\\
\sum_{j,i\in S_j}\hat q_{ji} + TQ_i=\sum_{k\in S_i}\hat
q_{ik},\forall i \in M-D,
\end{array}
\right.
\]
where the second constraint is the energy constraint and the third
constraint is for flow conservation. Notice that
(\ref{linear_prog}) has a linear objective function and can also
be written in a max-min form.

In practice, it is difficult to find $C_j^m$ since the complexity
will increase exponentially with the number of potential
collaborative/cooperative nodes in the set. One possible way to
simplify the calculation of the set $C_j^m$ is to select the
nearest neighbor for CB/CT. $C_j^m=1$, if node $j$'s nearest
neighbor can help node $j$ to reach node $m$. Obviously, this
simplification is suboptimal for (\ref{linear_prog}).

Next, if the information rate is random, each node dynamically
updates its cost according to its remaining energy and with
consideration of CB/CT. Some heuristic algorithms can be proposed
to update the link cost dynamically. Here the initial energy is
$E_i$. Denote the current remaining energy as $\underline{E}_i$.
We define the cost for node $i$ to communicate with node $j$ as

\begin{equation}\label{link_cost}
\mbox{cost}_{ij}=\left(\frac{E_i}{\underline{E}_i}\right)^{\beta_1}+
\sum _{l\in C_i^j} \left(
\frac{E_l}{\underline{E}_l}\right)^{\beta_2},
\end{equation}
where $\beta_1$ and $\beta_2$ are positive constants. Their values
affect the routing algorithms to allocate the packets between the
energy-sufficient and energy-depleting nodes, and between the
direct transmission and CB/CT. Notice that if $\beta_2=0$ in
(\ref{link_cost}), the cost function degrades to the normalized
residual energy cost function in \cite{Tassiulas}.

We assume that each node periodically broadcasts a HELLO packet to
its neighbors to update the topology information. Therefore, the
minimization problem in (\ref{Prob_def}) can be solved by applying
any distributed shortest path routing algorithm such as the
Bellman-Ford algorithm \cite{bib:Gallager}. There are two
differences to the traditional shortest path problem. First,
beyond direct transmission links, there are additional links
constructed by CT/CB. Second, when CT/CB links are used, the
energy is consumed not only in the transmission node but also in
the collaborative/cooperative nodes.

\section{Simulation Results}\label{sec:simulation}

We assume the sensor nodes and one sink are randomly located
within a square of size $\mathbb{L}\times \mathbb{L}$. Each node
has power of 10dbm, and the noise level is -70dbm. The propagation
loss factor is $4$. The minimal link SNR $\gamma_0$ is 10dB. The
initial energy of all nodes is assumed to be unit and the average
information rates for all nodes are 1.

\begin{figure}[htbp]
\begin{center}
    \epsfig{file=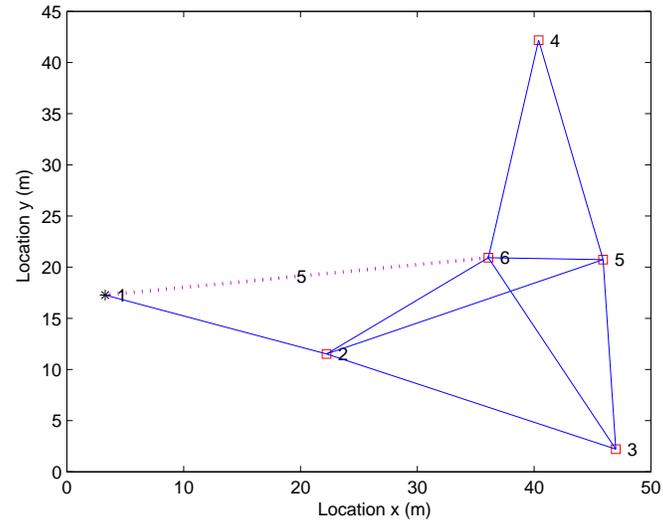,width=100truemm}
\end{center}
\caption{Snapshot of CB/CT Routing}\label{sim_example}
\end{figure}

In Figure \ref{sim_example}, we show a snapshot of a network of
$5$ sensor nodes and a sink with $\mathbb{L}=50$m. Here node $1$
is the sink. The solid lines are the direct transmission links,
and the dotted line from node $6$ to the sink is the CB/CT link
with the help of node $5$. For the traditional packet-forwarding
scheme, the best flow is

\begin{equation}
\hat q_{ij}=\left[
\begin{array}{cccccc}
  0 & 0 & 0 & 0 & 0 & 0 \\
  1 & 0 & 0 & 0 & 0 & 0 \\
  0 & 0.2 & 0 & 0 & 0 & 0 \\
  0 & 0 & 0 & 0 & 0.1 & 0.1 \\
  0 & 0.3 & 0 & 0 & 0 & 0 \\
  0 & 0.3 & 0 & 0 & 0 & 0 \\
\end{array}
\right]
\end{equation}
with the energy consumed for all nodes given by $[0,
1.0,0.2,0.2,0.3,0.3]$. Because node $2$ is the only node that can
communicate with the sink, the best network lifetime is limited to
$0.2$ before node $2$ runs out of battery.

With a new link from node $6$ to node $1$ with CB/CT, the best
flow is
\begin{equation}
\hat q_{ij}=\left[
\begin{array}{cccccc}
  0 & 0 & 0 & 0 & 0 & 0 \\
  1 & 0 & 0 & 0 & 0 & 0 \\
  0 & 0.321 & 0 & 0 & 0 & 0.012 \\
  0 & 0 & 0 & 0 & 0 & 0.333 \\
  0 & 0.23 & 0 & 0 & 0 & 0.103 \\
  0.667 & 0.115 & 0 & 0 & 0 & 0 \\
\end{array}
\right]
\end{equation}
with the energy consumed for all nodes given by $[0, 1.000, 0.333,
0.333, 1.0000, 0.782]$. Here some flow can be sent to the sink via
node $6$ instead of node $2$. Because of CB/CT, node $5$ has to
consume its own energy. The lifetime becomes $0.333$ which is a
67\% improvement over packet forwarding.

\begin{figure}[htbp]
\begin{center}
    \epsfig{file=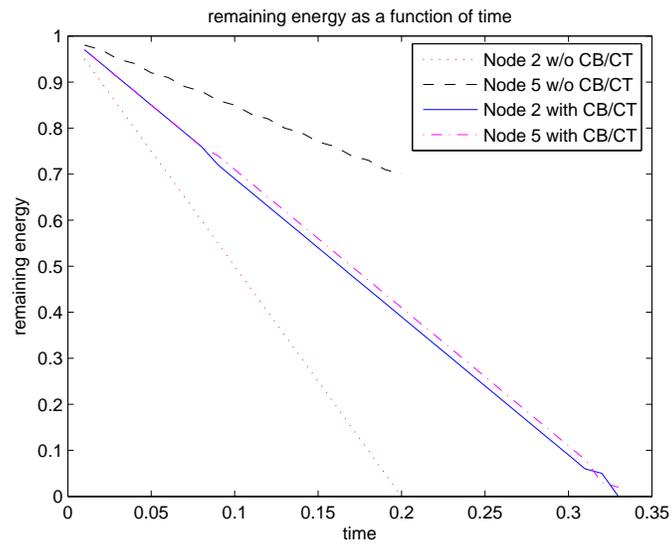,width=100truemm}
\end{center}
\caption{Remaining Energy over Time with and without
CB/CT}\label{energy_time}
\end{figure}

In Figure \ref{energy_time}, we show the dynamic behavior for the
algorithm using (\ref{link_cost}). Here $\beta=\beta_1=\beta_2=1$.
We show the remaining energy of node $2$ and node $5$ over time
with and without CB/CT, respectively. We can see that without
CB/CT, node $2$ runs out of battery energy at time $0.2$. Since
node $2$ is the only node that is able to connect to the sink, the
whole network dies at time $0.2$, even though node $5$ still has
more than 70\% of its energy left. With CB/CT, node $5$ can help
the link from node $6$ to the sink, so as to relieve the
packet-forwarding burden of node $2$. As a result, node $2$ can
extend its lifetime to $0.333$. In this example, we show that the
dynamic algorithm using (\ref{link_cost}) has the same performance
as that of the linear programming solution using
(\ref{linear_prog}). If we increase $\beta$, the curves of node
$2$ and node $5$ with CB/CT coincide with each other.

\begin{figure}[htbp]
\begin{center}
    \epsfig{file=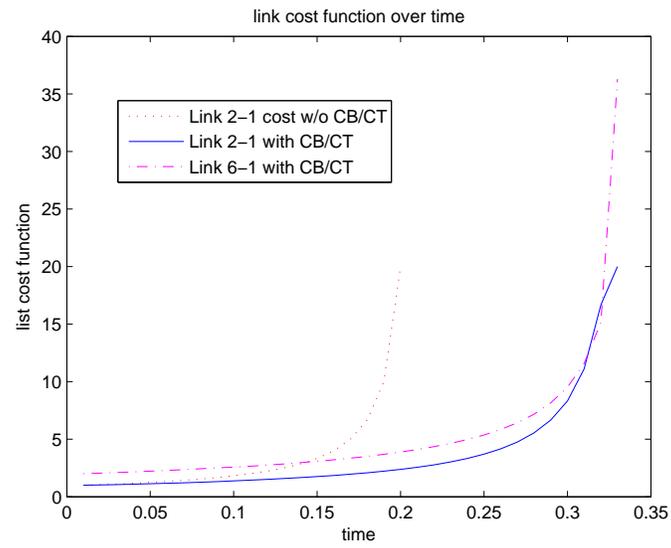,width=100truemm}
\end{center}
\caption{Link Cost over Time with and w/o CB/CT}\label{cost_time}
\end{figure}

In Figure \ref{cost_time}, we show the link cost changes over
time. We can see that the cost for each link increases almost
linearly with the increase of time or equivalently with the
decrease of the remaining energy at the beginning. When the nodes'
energy is critical, the price will increase quickly. The function
in (\ref{link_cost}) can be viewed as a boundary function
\cite{Convex} for the constrained optimization problem. If the
optimization is a minimization, the boundary function increases
the objective function significantly when the constraints are not
satisfied. Here the equivalent constraint is the non-negative
remaining energy. Similar to the boundary function, the cost
functions in (\ref{link_cost}) prevent the routing algorithm from
using the energy-depleting nodes. For different values of $\beta$,
the link costs have different sensitivities to remaining energy.

\begin{figure}[htbp]
\begin{center}
    \epsfig{file=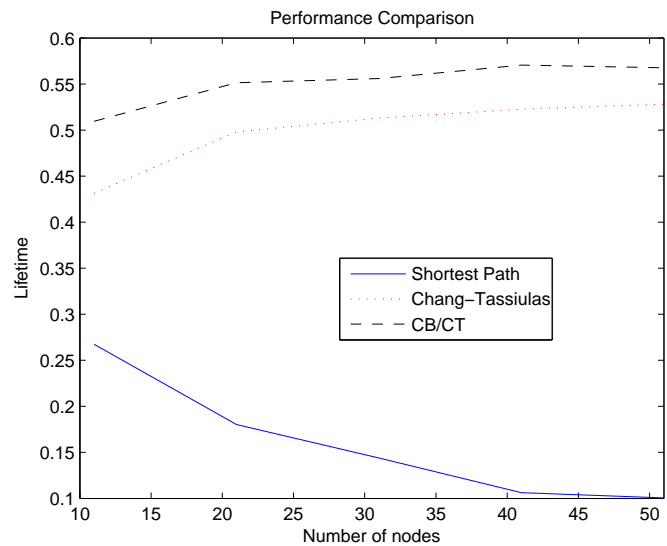,width=100truemm}
\end{center}
\caption{Lifetime Improvement Comparison} \label{life_no_user}
\end{figure}

In Figure \ref{life_no_user}, we compare the performance of three
algorithms: the shortest path, the algorithm in \cite{Tassiulas},
and the proposed CB/CT algorithm. Here $\mathbb{L}=100$m. As the
number of nodes increases, the performance of the shortest path
algorithm decreases. This decrease happens because more nodes need
packet forwarding by the nodes near the sink. Consequently, the
nodes near the sink die more quickly and the network lifetimes are
thus shorter. Compared with the algorithm in \cite{Tassiulas}, the
proposed schemes have about 10\% performance improvement. This
improvement is achieved because of the alternative routes to the
sink that can be established by CB/CT. With the increase in the
number of nodes, the performance of the algorithm in
\cite{Tassiulas} and the proposed algorithm can be slightly
improved, because there are more choices for the sensor nodes to
connect to the sink.

\begin{figure}[htbp]
\begin{center}
    \epsfig{file=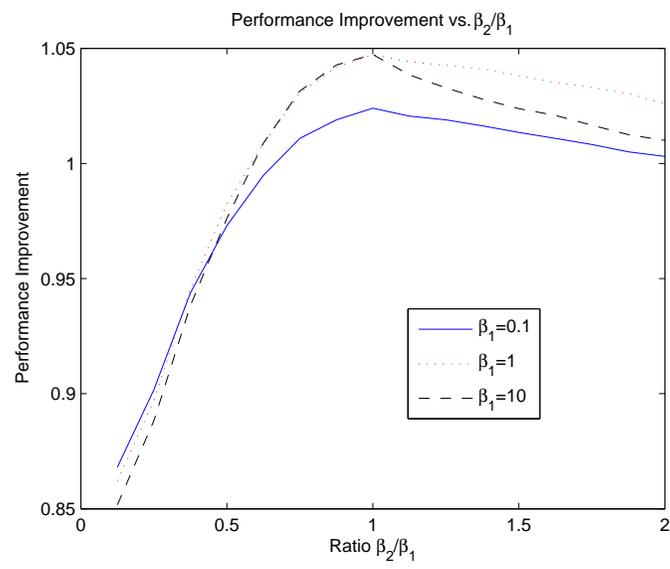,width=100truemm}
\end{center}
\caption{Lifetime Improvement vs. Ratio $\beta_2/\beta_1$}
\label{life_ratio}
\end{figure}

In Figure \ref{life_ratio}, we show the lifetime improvement over
the pure packet-forwarding scheme as a ratio of $\beta_2/\beta_1$
for $\beta_1=0.1$, $1$, and $10$, respectively. Recall that
$\beta_1$ and $\beta_2$ represent the willingness of the nodes to
participate in packet forwarding or in CB/CT. When
$\beta_2<\beta_1$, the nodes prefer CB/CT. Because CB/CT wastes
overall energy compared to packet forwarding, the performance of
the proposed scheme can be even worse than that of the pure
packet-forwarding scheme. When $\beta_2>\beta_1$, the nodes prefer
packet forwarding, and the performance degrades gradually with the
increase of the ratio. When $\beta_2=\beta_1$, the performance
improvement is maximized. This is because the nodes have no
difference between CB/CT and packet forwarding for the cost
function. (Notice that if the overhead for signalling is
considered, the above claim may no longer be not valid.) When
$\beta_1$ is small, the performance improvement is also small.
This is because the costs of the energy-depleting nodes are not
significantly higher than those of the energy-redundant nodes. As
a result, the routing algorithms still select the energy-depleting
nodes and the network lifetime is short. When $\beta_1=\beta_2$
and they are sufficiently large, the performance differences from
$\beta_1=1$ and $\beta_1=10$ are trivial. When $\beta_2$ is
different from $\beta_1$, the performance degrades faster for
larger $\beta_1$. This fact is because the values of the cost
functions in (\ref{link_cost}) are very sensitive to the values of
$\beta_1$ and $\beta_2$ when they are large.

\section{Conclusions}\label{sec:conclusion}

In this paper, we have studied the impact of CB/CT on the design
of higher level routing protocols. Specifically, using CB/CT, we
have proposed a new idea to bypass the energy-depleting nodes and
communicate directly with sinks or faraway nodes, so as to improve
the lifetimes of wireless sensor networks. First, we have
developed a closed-form analysis of the effectiveness of CB/CT to
enhance a wireless link. With this enhancement, the new routes can
be constructed by CB/CT, so that the information flow can avoid
the energy-depleting nodes. We have proposed static and dynamic
routing algorithms. In addition, we have investigated the dynamic
behavior of the proposed algorithms and studied the preferences
between packet forwarding and CB/CT. From the analytical and
simulation results, we have seen that the proposed algorithms can
reduce the payloads of energy-depleting nodes by about 90\% in a
2D disk case and increase lifetimes about 10\% in general
networks, compared with existing scheme techniques.

\bibliographystyle{IEEE}

\end{document}